\documentclass[twocolumn,showpacs,amsmath,amssymb,floatfix,pre]{revtex4}

\usepackage{graphicx}

\newcommand{\be}{\begin{equation}}
\newcommand{\ee}{\end{equation}}

\newcommand{\bea}{\begin{eqnarray}}
\newcommand{\eea}{\end{eqnarray}}

\begin{document}

\title{Molecular correlations and solvation in simple fluids}
\author{Marco A. A. Barbosa}
\altaffiliation{Current address: Faculdade UnB Planaltina, Universidade de Bras\'ilia, Bras\'ilia-DF, Brazil}
\affiliation{Department of Chemistry, Baker Laboratory, Cornell University,  \\
Ithaca, New York  14853-1301}
\author{B. Widom}
\thanks{Corresponding author}
\email[electronic address: ]{bw24@cornell.edu}
\affiliation{Department of Chemistry, Baker Laboratory, Cornell University,  \\
Ithaca, New York  14853-1301}

\begin{abstract}
 We study the molecular correlations in a lattice model of a solution of a low-solubility solute, with emphasis on how the thermodynamics is reflected in the correlation functions.  The model is treated in Bethe-Guggenheim approximation, which is exact on a Bethe lattice (Cayley tree).  The solution properties are obtained in the limit of infinite dilution of the solute.  With $h_{11}(r)$, $h_{12}(r)$, and $h_{22}(r)$ the three pair correlation functions as functions of the separation $r$ (subscripts $1$ and $2$ referring to solvent and solute, respectively), we find for $r \geq 2$ lattice steps that $h_{22}(r)/h_{12}(r) \equiv  h_{12}(r)/h_{11}(r)$.  This illustrates a general theorem that holds in the asymptotic limit of infinite $r$.  The three correlation functions share a common exponential decay length (correlation length), but when the solubility of the solute is low the amplitude of the decay of $h_{22}(r)$ is much greater than that of $h_{12}(r)$, which in turn is much greater than that of $h_{11}(r)$. As a consequence the amplitude of the decay of $h_{22}(r)$ is enormously greater than that of $h_{11}(r)$. The effective solute-solute attraction then remains discernible at distances at which the solvent molecules are essentially no longer correlated, as found in similar circumstances in an earlier model.  The second osmotic virial coefficient is large and negative, as expected.  We find that the solvent-mediated part $W(r)$ of the potential of mean force between solutes, evaluated at contact, $r=1$, is related in this model to the Gibbs free energy of solvation at fixed pressure, $\Delta G_p^*$, by $(Z/2) W(1) + \Delta G_p^* \equiv p v_0$, where $Z$ is the coordination number of the lattice, $p$ the pressure, and $v_0$ the volume of the cell associated with each lattice site.  A large, positive $\Delta G_p^*$ associated with the low solubility is thus reflected in a strong attraction (large negative $W$ at contact), which is the major contributor to the second osmotic virial coefficient.  In this model, the low solubility (large positive $\Delta G_p^*$) is due partly to an unfavorable enthalpy of solvation and partly to an unfavorable solvation entropy, unlike in the hydrophobic effect, where the enthalpy of solvation itself favors high solubility, but is overweighed by the unfavorable solvation entropy.
\end{abstract}

\date{\today}
\maketitle 

\section{Introduction}

When a solute is only sparingly soluble in a solvent, pairs of solute molecules effectively attract.  In the extreme, the solutes aggregate and precipitate as a separate phase in equilibrium with a very dilute solution of that solute in the solvent.  We propose here to obtain quantitative relations between the solute's low innate solubility and the potential of mean force between a pair of solute molecules, the latter describing the effective attraction between them.  Of particular interest is the solvent-mediated part of the potential of mean force; \textit{i.e.}, the part remaining after the direct solute-solute interaction is subtracted.

We study this in a lattice model on a Bethe lattice (Cayley tree), for which the Bethe-Guggenheim ``quasichemical'' approximation~\cite{rushbrooke:statmech,hill:statmech} becomes exact. The Bethe-Guggenheim and related approximations have been used before for lattice models of water and hydrophobic solvation~\cite{bell70:jphysa1,bell70:jphysa2,besseling97,eads02,widom03:pccp,henriques06} and the approximation is used extensively in continuum models of solutions as well~\cite{beck:potential,rogers08}.  Models of water and aqueous solutions on other hierarchical lattices have also been extensively studied~\cite{barbosa08,buzano03:jcp}.  The equivalent of the present model but with emphasis on its properties in the neighborhood of its critical-point manifold rather than in the thermodynamic states of interest here has also been analyzed previously~\cite{wheeler70:jcp}.  Here we are primarily interested in the limit of infinite dilution of the solute in the dense liquid solvent.

In the following section we review and collect the properties of the pure solvent on the Bethe lattice, which we will need subsequently.  In Section~\ref{solution} we find the thermodynamic properties and correlation functions of the solution in the limit of an infinitely dilute solute.  These are the main results in the paper, and are illustrated numerically with a representative set of the model's parameters.  Section~\ref{limits} re-expresses the results in two limits, both helping to illuminate the general formulas.  One is the mean-field limit, in which the coordination number $Z$ of the lattice becomes infinite while the solvent-solvent interaction energy $-\epsilon_{11}$ goes to $0$, with $Z\epsilon_{11}$ fixed.  The other is the low-temperature limit, $T\rightarrow 0$, while the high-density liquid solvent remains in coexistence with its low-density equilibrium vapor.  The final Section~\ref{summary} is a brief summary of our results.

\section{\label{solvent}Solvent properties}

Here we summarize the properties of the one-component system, without solute, which will then be the properties of the pure solvent we shall need in the later parts of the paper.  The formulas here are either well known or readily derivable and they establish our notation.

Our Bethe lattice is of coordination number $Z$.  Associated with each site is a cell of volume $v_0$~\cite{wheeler70:jcp}.  Only singly occupied and empty cells are allowed.  Except for the exclusion of multiple occupancy, the only interactions are between molecules in neighboring cells, which interact with energy $-\epsilon_{11}$  ($\epsilon_{11}>0$).  With $\rho$ the density, $T$ the temperature, $p$ the pressure, and $k$ Boltzmann's constant, we introduce the abbreviations
\be
y = \rho v_0, \;\;\;t = kT/\epsilon_{11},  \;\;\; \pi = p v_0/\epsilon_{11},  \;\;\;c = e^\frac{1}{t}. \label{eq:def}
\ee
The $\pi$, $y$, $t$ equation of state on the Bethe lattice (``quasichemical'' approximation) is then
\be
\frac{\pi}{t} = \ln \left [ (1-y)^{\frac{1}{2}Z-1} (1+\alpha)^{\frac{1}{2}Z}  \right ],\label{eq:pi}
\ee
where
\be
\alpha = \frac{2y-1+Q}{2(1-y)c}\label{eq:alpha}
\ee
with
\be
Q=\sqrt{1+4(c-1)(1-y)y}. \label{eq:Q}
\ee

We define a dimensionless compressibility $\chi$ and coefficient of thermal expansion $\theta$ by
\be
\chi = \left( \frac{\partial \ln y }{ \partial \ln \pi} \right)_t, \;\;\; \theta = -\left( \frac{\partial \ln y }{ \partial \ln t} \right)_\pi,
\label{eq:compress}
\ee
these being, respectively, the fractional rate of increase of density with increasing pressure at fixed temperature and the fractional rate of decrease of density with increasing temperature at fixed pressure.  In the high-density states of interest here, these will both be very small ($\ll 1$), as in real liquids.

The equation of the liquid-vapor coexistence curve in the $y$,$t$ plane is
\be
\frac{1}{t} = 2 \ln \left[ \frac{\kappa^Z -1}{ \kappa( \kappa^{Z-2} -1)} \right ],  \;\;\; \kappa \equiv \left( \frac{1}{y} -1  \right)^{1/Z} \texttt{      (coex.)}. \label{eq:coex}
\ee
The equation of the spinodal is
\be
Q = \frac{Z}{Z-2} \texttt{      (spinodal)}. \label{eq:spinodal}
\ee
The spinodal lies within the coexistence curve in the $y$,$t$ plane and is tangent to it at the critical point.  The critical point $y_c$, $t_c$ is at
\be
y_c = \frac{1}{2},  \;\;\; \frac{1}{t_c} = 2 \ln \frac{Z}{Z-2}.\label{eq:tc}
\ee

We measure distances $r$ between cells as the number of steps on the underlying lattice.  This defines it uniquely because there are no closed loops on a Bethe lattice (Cayley tree).  The pair correlation function $h_{11}(r)$ is
\be
h_{11}(r) = \left( \frac{1}{y} -1  \right)e^{-r/\xi}  \;\;\; (r \geq 1), \;\;\; h_{11}(0) = -1, \label{eq:h11}
\ee
where the correlation length (exponential decay length) $\xi$ is
\be
\xi = \left [ \ln \frac{Q+1}{Q-1} \right] ^{-1}. \label{eq:xi}
\ee
When, as in~(\ref{eq:h11}) and~(\ref{eq:xi}) with $Q$ as in (\ref{eq:Q}), the thermodynamic state is specified by the variables $y$ and $t$, the correlation function is independent of coordination number $Z$ and so is the same as when $Z=2$ (the linear chain).  That is a reflection of the uniqueness of the path between any two sites on the lattice.  But it is no longer true when some other variable such as $\pi$ in Eq.~(\ref{eq:pi}), or the activity, replaces either or both of $y$ and $t$; then $h_{11}(r)$ does depend on $Z$.  Likewise, when the state is specified to be at the coexistence curve, $h_{11}(r)$ again depends on $Z$ via~(\ref{eq:coex}).  Indeed, there is no coexistence curve unless $Z > 2$.

It is a peculiarity of the Bethe lattice that the correlation length $\xi$ in~(\ref{eq:xi}) remains finite at the critical point.  Nevertheless, the sum of $h_{11}(r)$ over all lattice sites diverges there, along with the compressibility, in accord with the Ornstein-Zernike theorem.  That is because the number $N_r$ of sites distant $r$ from any central site diverges exponentially rapidly as $r\rightarrow \infty$:
\be
N_r = Z(Z-1)^{r-1} \;\;\; (r\geq 1), \;\;\; N_0 =1.\label{eq:Nr}
\ee
Thus, as may be verified from~(\ref{eq:pi})-(\ref{eq:compress}) and~(\ref{eq:tc})-(\ref{eq:Nr}),
\be
t \chi/\pi \equiv 1/y+\sum_{r=0}^\infty N_r h_{11}(r), \label{eq:oz}
\ee
as required by the Ornstein-Zernike theorem. Both sides diverge at the critical point given by~(\ref{eq:tc}).

\section{\label{solution}Solution with infinitely dilute solute}

We define the solubility $\Sigma$ of the solute (species 2) to be the dimensionless ratio of its number density $\rho_2$ in the solution to its activity $z_2$, in the limit of infinite dilution, with the activity defined to become equal to its number density in an ideal gas; thus,
\be
\Sigma \equiv \rho_2/z_2 \label{eq:def:sigma}. 
\ee
Except for a factor of $kT$, this is also the Henry's-law constant, and it is also the Ostwald absorption coefficient when the vapor in equilibrium with the solution is an ideal gas. Low solubility means $\Sigma \ll 1$.

By the potential-distribution theorem (particle-insertion theorem), $\Sigma$ is then the product of the probability $1-y$ that a trial solute molecule inserted into the pure solvent has found an empty cell, multiplying the conditional average $ \langle e^{-\Psi/kT} \rangle $, with $\Psi$ the energy of interaction of the inserted solute with its (hypothetically undisturbed) neighboring solvent molecules.  On the Bethe lattice of coordination number $Z$, the probability $f_j$ that an empty cell in the pure solvent is neighbored by exactly $j$ occupied cells is~\cite{wheeler70:jcp}
\be
f_j = \frac{Z!}{j!(Z-j)!} \frac{\alpha^j}{(1+\alpha)^Z},
\ee
with $\alpha$ as in~(\ref{eq:alpha}).  Thus,
\bea
\Sigma & = & (1-y) \sum_{j=0}^Z f_j e^{j\epsilon_{12}/kT} \nonumber  \\
       & = & (1-y) \left( \frac{1+\alpha c^a}{ 1+\alpha } \right)^Z, \label{eq:sigma}
\eea
where $-\epsilon_{12}$ is the interaction energy between solute and solvent molecules in neighboring cells, where $c$ is as in~(\ref{eq:def}), and where
\be
a = \epsilon_{12}/\epsilon_{11}.\label{eq:defa}
\ee
This gives $\Sigma$ as a function of temperature $t$ and solvent density $y$. In the dense liquid solvent at given $t$ it is a very sensitive function of $y$ but insensitive to pressure $\pi$ because of the solvent's low compressibility. We shall later wish to know the rate at which $\Sigma$ changes with temperature at fixed pressure, which may be obtained from (\ref{eq:pi})-(\ref{eq:compress}) and (\ref{eq:sigma}) by the identity
\be
\left (\frac{\partial \ln \Sigma  }{ \partial \ln t } \right )_\pi \equiv \left (\frac{\partial \ln \Sigma  }{ \partial \ln t } \right )_y - \left (\frac{\partial \ln \Sigma  }{ \partial \ln y } \right )_t \theta.
\ee

Along with $h_{11}(r)$ in~(\ref{eq:h11}) and~(\ref{eq:xi}) we require also the pair correlation functions $h_{12}(r)[\equiv h_{21}(r)]$ between solute-solvent pairs and $h_{22}(r)$ between pairs of solutes, again in the limit of infinite dilution.  The results are given here, with the detailed derivations in Appendix A.

As shown in the appendix,
\be
\frac{h_{12}(r)}{h_{11}(r)} \equiv D \;\;\; (r \geq 1), \;\;\;\;\; \frac{h_{22}(r)}{h_{12}(r)} \equiv D \;\;\; (r \geq 2), \label{eq:relh}
\ee
where $D = h_{12}(1)/h_{11}(1)$ is a constant, independent of $r$. As also shown in the appendix,
\be
h_{12}(1) = -1+\frac{1/y}{1+1/(c^a\alpha)}\label{eq:h12c}
\ee
with $c$ and $\alpha$ as defined in~(\ref{eq:def}), (\ref{eq:alpha}), and (\ref{eq:Q}); while from~(\ref{eq:h11}) and (\ref{eq:xi}),
\be
h_{11}(1) =\left ( \frac{1}{y}-1  \right) \frac{Q-1}{Q+1}. \label{eq:h11c}
\ee
From~(\ref{eq:h12c}) and~(\ref{eq:h11c}) we then have  the constant $D$.
In the meantime, $h_{22}(1)$ contains the solute-solute interaction energy, $-\epsilon_{22}$. We define
\be
b=\epsilon_{22}/\epsilon_{11}, \label{eq:defb}
\ee
analagously to the definition of $a$ in~(\ref{eq:defa}). Then as further shown in the appendix,
\be
h_{22}(1)= -1+\left[ 1+ h_{11}(1) \right ] c^{b-1} \left(  \frac{1+c\alpha}{ 1+c^a\alpha }  \right )^2. \label{eq:h22c} 
\ee
We also have $h_{12}(0)=h_{22}(0)=-1$. From~(\ref{eq:h11}), (\ref{eq:xi}), and~(\ref{eq:relh})-(\ref{eq:h22c}) we now have all three correlation functions $h_{11}(r)$, $h_{12}(r)$, $h_{22}(r)$ for all $r$.

These results were derived both by the evaluation of the appropriate constrained partition functions and by transfer-matrix methods.  The two derivations are equally lengthy; the  appendix outlines the derivation via transfer matrices.

Like $h_{11}(r)$ in~(\ref{eq:h11}) and~(\ref{eq:xi}), here, too, $h_{12}(r)$ and $h_{22}(r)$ are independent of $Z$ when the thermodynamic state of the system is specified by the temperature $t$ and solvent density $y$, and so are the same as for the linear chain, $Z=2$.  
 
The analogs of the Ornstein-Zernike relation for a two-component fluid are the Kirkwood-Buff relations~\cite{ben-naim77:jcp,kirkwood51,zimm53,ben-naim:water}. In the limit in which the solute $2$ is infinitely dilute in solvent $1$, this requires for the present model that
\be
t\chi/\pi = v_2/v_0 + \sum_{r=0}^{\infty} N_r h_{12}(r),\label{eq:kb}
\ee
where $v_2$ is the partial molecular volume of the solute at infinite dilution in the solvent. This is the analog of the Ornstein-Zernike relation~(\ref{eq:oz}) with $h_{12}(r)$ in place of $h_{11}(r)$ and $v_2/v_0$ in place of $v_1/v_0=1/y$. This $v_2/v_0$ is in turn related by thermodynamic identity to the rate of change of solubility $\Sigma$ with the density $y$ of the solvent~\cite{widom07:jpcc},
\be
\frac{v_2}{v_0}=-t \frac{\chi}{\pi}\left [ \left ( \frac{\partial \ln \Sigma}{ \partial \ln y } \right)_t - 1 \right ],
\ee
so from~(\ref{eq:kb}), it is required that
\be
\sum_{r=0}^{\infty} N_r h_{12}(r) = t \frac{\chi}{\pi} \left ( \frac{\partial \ln \Sigma}{ \partial \ln y } \right)_t. \label{eq:oz-kb}
\ee
One may verify from~(\ref{eq:oz}), (\ref{eq:sigma}), and (\ref{eq:relh})-(\ref{eq:h11c}), with $h_{11}(0)=h_{12}(0)=-1$, that the required~(\ref{eq:oz-kb}) is satisfied. This is an important consistency test.

From~(\ref{eq:relh}) we have $h_{22}(r)/h_{12}(r) \equiv h_{12}(r)/h_{11}(r) \;\; ( r\geq 2)$. This may be recognized to be a special case of a general principle definitively discussed in an important paper by Evans \textit{et al.}~\cite{evans94:jcp}, who show that a similar relation holds asymptotically as $r\rightarrow \infty$ for any mixture, not only in the limit in which some of the components are present at infinite dilution.  In the present context it holds not only for $r\rightarrow \infty$ but for all $r \geq 2$.   One sees also in~(\ref{eq:relh}) that $h_{12}(r)$ and $h_{22}(r)$ have the same exponential decay length $\xi$ as in the solvent, given in~(\ref{eq:xi}).  That is because the correlations between a given solvent molecule and a molecule of an infinitely dilute solute, or between two such solute molecules, propagate through the solvent and so must have the range of the solvent-solvent correlations themselves.

In the low-solubility states that are our primary interest, the common ratio $D$ in~(\ref{eq:relh}), evaluated as $D=h_{12}(1)/h_{11}(1)$ via~(\ref{eq:h12c}) and~(\ref{eq:h11c}), proves to be very large (and negative), so the amplitude of $h_{22}(r)$ is then much greater than that of $h_{12}(r)$, which in turn is much greater than that of $h_{11}(r)$.  As a consequence, the amplitude of $h_{22}(r)$ is enormously greater than that of $h_{11}(r)$.  This property was earlier observed in a one-dimensional model of molecules interacting with square-well potentials~\cite{widom08:mp,widom07:jpcc}.  It is likely to be generally true in solutions of solutes of low solubility, as a reflection of the effective, solvent-mediated attraction between them.

The effective attraction between solute molecules is manifested also in the second osmotic virial coefficient, $B$.  In a continuum model, with solute species $2$ at infinite dilution, this would be given by $-1/2$ of the integral of $h_{22}(r)$ over all space.  In the present lattice model this is
\be
\frac{B}{v_0}=-\frac{1}{2}\sum_{r=0}^\infty N_r h_{22}(r)\label{eq:def:soc}
\ee
with $N_r$ in~(\ref{eq:Nr}).  With $h_{22}(r)$ as above, and from the earlier formulas, the summation may be done explicitly and the result expressed as
\be
B = B_0+B_1+B_2,
\ee
where these are the contributions from $r=0$, $r=1$, and $r \geq 2$, respectively:
\begin{widetext}
\begin{subequations}
\bea
\frac{B_0}{v_0} & = & \frac{1}{2}, \\
\frac{B_1}{v_0} & = & - \frac{1}{2}Z \left \{ -1+ \left[ 1+ \left ( \frac{1}{y} - 1 \right ) \frac{Q-1}{Q+1}  \right ] c^{b-1} \left ( \frac{1+c\alpha}{ 1+c^a\alpha} \right )^2  \right  \} , \\
\frac{B_2}{v_0} & = & -\frac{1}{2}Z(Z-1)D^2 \left ( \frac{1}{y} - 1 \right ) \left (\frac{Q-1}{Q+1} \right )^2 \frac{1}{1-(Z-1) \frac{Q-1}{Q+1}}  .
\eea
\end{subequations} 
\end{widetext}
In the states of interest here, $B/v_0$ will prove to be large and negative, with the major contribution coming from $B_1/v_0$; \textit{i.e.}, from the solute pair in contact.  If we think of the negative $B$ as arising from the ``dimerization''~\cite{tucker79:jpc,rossky80:jpc,chipot96:jcc,liu98:jpcb,coskuner06:zpc,coskuner07:zpc},
\be
2\texttt{ solute monomers}\; \rightleftharpoons \; \texttt{solute dimer}, \nonumber
\ee   
with ``monomer'' and ``dimer'' both dilute in the solvent, then $-B$ is the ``equilibrium constant'' for this dimerization.

Let $\Delta G_p$, $\Delta S_p$, and $\Delta H_p$ be the net changes in free energy, entropy, and enthalpy when a solute molecule is transferred at fixed $p$ and $T$ from an ideal gas phase where its number density is $\rho_2(\texttt{i.g.})$ to the solution where its number density is $\rho_2(\texttt{soln.})$.  We define the constant-pressure solvation free energy and entropy $\Delta G^*_p$ and $\Delta S^*_p$ by 
$\Delta G^*_p = \Delta G_p - kT \ln \left [ \rho_2(\texttt{soln.})/\rho_2(\texttt{i.g.})   \right ] $ 
and 
$\Delta S^*_p = \Delta S_p + k\ln \left [ \rho_2(\texttt{soln.})/\rho_2(\texttt{i.g.})   \right ] $, while $\Delta H_p$ itself is the constant-pressure solvation enthalpy~\cite{widom03:pccp}.  These are related by $\Delta G^*_p = \Delta H_p -T \Delta S^*_p$.  When the ideal gas is in equilibrium with the solution, $\rho_2(\texttt{i.g.})$ is the same as the common activity $z_2$ in the two phases, while                            $\Delta G_p=0$, so
\be
\Delta G^*_p=-kT\ln \Sigma \label{eq:def:dG}
\ee
with the solubility $\Sigma$ as defined in~(\ref{eq:def:sigma}). In the limit of infinite dilution, $\Sigma$ is independent of the separate values of $\rho_2$ and $z_2$; it depends on the nature of the solute and solvent and on their interactions but is otherwise a function only of the thermodynamic state of the solvent (Henry's law).  The solvation entropy $\Delta S^*_p$ and enthalpy $\Delta H_p$ as defined above are then given by~\cite{widom03:pccp}
\begin{subequations}
\bea
 \frac{\Delta S^*_p}{k} & = & \left [ \frac{\partial}{\partial T}(T \ln \Sigma )  \right ]_p + \theta -1, \\
\frac{\Delta H_p}{kT} & = & T \left ( \frac{\partial \ln \Sigma}{\partial T}  \right )_p + \theta -1
\eea
\end{subequations}
where $\theta$ is the dimensionless coefficient of thermal expansion defined in~(\ref{eq:compress}) and the $-1$ is the analogous quantity for the ideal gas.

\begin{figure}
\begin{center}
\includegraphics[scale=0.65]{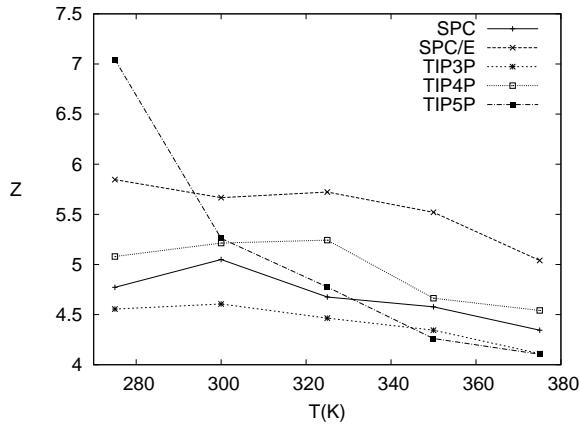}
\caption{Coordination number $Z$ from fits of~(\ref{eq:lrel}) with right-hand side $0$ to simulation data of Paschek~\cite{paschek04:jcp} for Xe in various water models as solvent.}
\label{fig:xenon}
\end{center}
\end{figure}

By~(\ref{eq:def:dG}), $\Delta G^*_p/kT$ is a measure of how low the solubility of the solute is. As remarked in the Introduction, a low solubility will be reflected in a strongly attractive potential of mean force between solutes.  We concentrate on the solvent-mediated part, $W(r)$, of this potential; \textit{i.e.}, on the part remaining after the direct solute-solute interaction is subtracted.  In particular, we wish to relate $\Delta G^*_p$ to $W(1)$, the contact value of $W(r)$, for we expect these to be closely correlated when the solubility is low, as found in an earlier lattice model of hydrophobic effects~\cite{widom03:pccp} and confirmed in simulations of realistic models of aqueous solutions of non-polar solutes~\cite{paschek04:jcp}. Indeed, as shown in Appendix B, we find these in the present Bethe-lattice model to be related by
\be
\frac{Z}{2} W(1)+\Delta G^*_p \equiv p v_0,\label{eq:lrel}
\ee
where, from~(\ref{eq:h11c}) and (\ref{eq:h22c}), but now without the factor $c^b$,
\be
e^{-W(1)/kT} = \left[ 1+ \left ( \frac{1}{y}-1  \right ) \frac{Q-1}{Q+1}   \right ] c^{-1} \left(  \frac{1+c\alpha}{ 1+c^a\alpha }  \right )^2. \label{eq:W1} 
\ee
In the states of most interest here, $pv_0$ is very much smaller than either $\Delta G^*_p$ or $-(Z/2)W(1)$, so the right-hand side of~(\ref{eq:lrel}) may often be replaced by $0$.  
When Paschek's simulation data~\cite{paschek04:jcp} for Xe in SPC, SPC/E, TIP3P, TIP4P, and TIP5P water, in the temperature range $275$ to $375$K, are fit by~(\ref{eq:lrel}) with right-hand side $0$, most of the resulting $Z$ lie between $4.2$ and $5.8$ (Fig.~\ref{fig:xenon}). Such values of the coordination number plausibly reflect the water structure.

We now illustrate numerically the formulas in this and the preceding section by evaluating them in a representative state in which the solute is of low solubility in the high-density liquid solvent that is in coexistence with its low-density vapor at a low temperature.  We recall the abbreviations in~(\ref{eq:def}), (\ref{eq:defa}), and~(\ref{eq:defb}).  We choose for illustration $Z=3$, $a=1/3$, $b=1/9$, and $t=1/5$.  (The relation $b=a^2$ is ``Berthelot's rule").  With this $Z$ the critical $t$ from~(\ref{eq:tc}) is $t_c=1/\ln 9 $, so we have chosen $t/t_c = \frac{1}{5}\ln 9 \approx 0.44$. In the states of interest, the choice $a<1/2$ proves to be essential for the solubility to be low. The reason for this will be seen clearly in the asymptotic limit derived in Section~\ref{limits}.

From the formulas in Section~\ref{solvent}, the liquid at the coexistence curve at this $t$ then has the properties
\bea
y   & = &   0.999268, \;\;\;\;\;    \pi = 0.000127 \nonumber \\
\xi &  = &   0.414  \nonumber \\
\chi & = & 6.16 \times 10^{-7}, \;\;\;  \theta = 6.09 \times 10^{-3}. \label{eq:solventnumbers}
\eea
This is a very high, nearly close-packed density and a very low pressure.  The exponential decay length is very short:  between $r$ and $r+1$ the correlation functions $h_{11}(r)$ and $h_{12}(r)$ for $r\geq 1$, and $h_{22}(r)$ for $r \geq 2$, decay by the factor $\exp (-1/\xi)=0.089$.  The compressibility and coefficient of thermal expansion are both very low, as in real liquids.

From the formulas in the present section, with values of the parameters as above,
\begin{align}
\Sigma &= 0.0865, \;\;\; \left ( \partial \ln \Sigma / \partial \ln t \right )_\pi = 2.74 \nonumber \\
D  & =  -269.2 \nonumber \\
B/v_0 &= -12.9\;\; \nonumber \\ 
( B_0/v_0 &=0.5,\; B_1/v_0=-11.9,\;B_2/v_0=-1.6 ) \nonumber \\
\Delta G^*_p/kT (&=-\ln \Sigma) = 2.447, \;\;\; W(1)/kT  = -1.631 \nonumber \\
\Delta S^*_p/k  &=  -0.698, \;\;\; \Delta H_p/kT = 1.749. \label{eq:solutenumbers}
\end{align} 
This, as anticipated because of the high solvent density and low temperature (and because $a<1/2$), is a low solubility $\Sigma$, in the range of solubilities of small non-polar solutes in water.  Also as anticipated, reflecting the low solubility, is a large negative $D$, which is the common ratio of $h_{22}(r)$ to $h_{12}(r)$ for $r\geq2$ and of $h_{12}(r)$ to $h_{11}(r)$ for $r \geq 1$.  As a consequence, $h_{22}(r)/h_{11}(r) = D^2 \approx 72,000$.  So great a ratio of amplitudes should be readily discernible by experiment or simulation if one could measure the correlations at large enough $r$.  For example, with these values of the model's parameters, $h_{11}(2)= 6\times 10^{-6}$ while $h_{22}(2)=0.4$; so, despite $h_{22}$ and $h_{11}$ having the same formal decay length, $h_{22}$ remains substantial long after $h_{11}$ has become negligible. That $D$ is negative reflects the effective repulsion between solute and solvent associated with the low solubility, while the solvent-solvent interactions and the effective, solvent-mediated solute-solute interactions are both attractive. The second osmotic virial coefficient is large and negative, again reflecting the effective attraction between solutes that accompany the low solubility.  As remarked earlier, the main contributor to the large negative $B$ is the effective (direct and solvent-mediated) solute-solute attraction at $r=1$. We note also that with the present $Z=3$ the relation between $W(1)$  and $\Delta G^*_p$ required by the identity~(\ref{eq:lrel}) is satisfied with $pv_0/kT \; (= \pi/t)$ negligibly small compared with both $-\frac{1}{2}Z W(1)/kT$ and $\Delta G^*_p/kT$.  

As a final remark, we observe that $(\partial \Sigma /\partial t )_\pi >0$,  $\Delta S^*_p <0$, and $\Delta H_p >0$.  
In these respects the low solubility here then contrasts with that in the hydrophobic effect, where $(\partial \Sigma /\partial T )_p < 0$,  $\Delta S^*_p <0$, and $\Delta H_p < 0$. There, the solubility decreases with increasing $T$, and the solubility is low because the solvation entropy is sufficiently unfavorable to overweigh a favorable solvation enthalpy, whereas in the present model the solubility increases with increasing $T$, and the solubility is low because the entropy and enthalpy of solvation are \textit{both} unfavorable.
In the present model, too, there are states in which $\left ( \partial \Sigma /\partial T \right )_p <0$, but they are not the low-solubility states of interest.

\section{\label{limits}Mean-field and low-temperature limits}

The formulas in Sections~\ref{solvent} and~\ref{solution} simplify greatly in the mean-field and low-temperature limits.

The mean-field limit is that in which $Z \rightarrow \infty$ and $\epsilon_{11} \rightarrow 0$ with the product $Z\epsilon_{11} \equiv \phi$ fixed.  In this limit, from $t$ in~(\ref{eq:def}) and from~(\ref{eq:tc}),
\be
\phi = 4kT_c,
\ee												
and then from~(\ref{eq:pi})-(\ref{eq:compress}),
\bea
pv_0 & = & - \frac{1}{2}\phi y^2+kT \ln \frac{1}{1-y} \label{eq:pi:mf} \\
\chi & = &  \frac{ \frac{T}{T_c}\ln \frac{1}{1-y} -2y^2  }{ \frac{T}{T_c}\frac{y}{1-y} -4y^2 } \\
\theta & = & -\frac{1}{y} \frac{ \ln \frac{1}{1-y}  }{ \; 4\frac{T}{T_c}y - \frac{1}{1-y} }.
\eea
The equations of the coexistence curve and spinodal, from~(\ref{eq:coex}) and~(\ref{eq:spinodal}), are
\bea
\frac{T}{T_c}  &=& \frac{4(y-\frac{1}{2})}{\ln \frac{y}{1-y}  } \;\;\;  \texttt{(coex.)}  \\
\frac{T}{T_c}  &=&  4y(1-y) \;\;\;  \texttt{(spinodal)}
\eea
and from~(\ref{eq:sigma}) the solubility becomes
\be
\Sigma = (1-y)e^{4 a y T_c/T}.\label{eq:sigma:mf}
\ee

Equation~(\ref{eq:sigma:mf}) may be understood from the potential-distribution formula.  The factor $(1-y)$ is the probability that the test solute molecule is in a previously empty cell, $Zy$ in the mean-field limit is the average number of neighboring cells occupied by solvent molecules, and the mean-field limit of the conditional average $\langle \exp(-\Psi/kT) \rangle$ is then $\exp(Zy\epsilon_{12}/kT) = \exp(Zya\epsilon_{11}/kT) = \exp(ya\phi/kT) = \exp(4yaT_c/T )$.  We see in this approximation that $\Sigma$ always decreases with increasing $T$ at fixed $y$ (although not at fixed $p$).

From~(\ref{eq:def:dG}) and~(\ref{eq:lrel}), in this mean-field limit,
\be
\frac{Z}{2} W(1) = -\frac{1}{2}y(y-2a)\phi, \;\;\; \Delta G^*_p = kT \ln \frac{1}{1-y} -ay\phi,
\ee
which, with~(\ref{eq:pi:mf}), is seen to satisfy the general~(\ref{eq:lrel}).

A more relevant and more interesting limit is that of low temperature, which is equivalently that of high density if we continue to specify the thermodynamic state of the liquid solvent as that at its coexistence curve, where it is in equilibrium with a low-density vapor. Then from~(\ref{eq:coex}), asymptotically, as $t$ and $1-y$ both approach $0$ on the coexistence curve, they are related by
\be
1-y \sim e^{-\frac{1}{2} Z/t}.
\ee
Now from the formulas of Section~\ref{solution}, in this limit, with $Z>2$ (otherwise there is no coexistence curve) and $a<1$ (the $1$,$2$ attraction weaker than the $1$,$1$ attraction),
\bea
\Sigma & \sim & e^{-Z(\frac{1}{2} -a)/t} \label{eq:sigma:lt} \\ 
D & \sim & -e^{(Z/2 -a)/t} \label{eq:D:lt} \\
\frac{\Delta G^*_p}{kT} &\sim& -\frac{Z}{2} \frac{W(1)}{kT} \sim \frac{Z(\frac{1}{2}-a) }{t} \label{eq:dG:lt}
\eea
and with $b=a^2$,
\begin{widetext}
\be
\frac{B_0}{v_0} = \frac{1}{2}, \;\;\; \frac{B_1}{v_0} \sim -\frac{1}{2}Ze^{(1-a)^2/t} , \;\;\;\frac{B_2}{v_0} \sim -\frac{1}{2}Z(Z-1)e^{[ -\frac{1}{2}Z +2(1-a)]/t}. \label{eq:B:lt}
\ee
\end{widetext}

From~(\ref{eq:sigma:lt}), it is now clear why $a<1/2$ was a condition for low solubility in the low-temperature, high-density solvent states of interest here.  Also, since $Z>2$, we see from~(\ref{eq:D:lt}) why $D$ is negative and of large magnitude, thus making the amplitude of $h_{22}(r)$ hugely greater, by the factor $D^2$, than that of $h_{11}(r)$ for all $r \geq 2$.
We note that the large $-D$ is correlated with the low solubility via $-D \sim \Sigma^{-(Z-2a)/Z(1-2a)}$, which, with $a<1/2$, is a negative power of $\Sigma$. We see from~(\ref{eq:B:lt}) why $B_1/v_0$ is negative and of large magnitude.  From~(\ref{eq:dG:lt}), we see that~(\ref{eq:lrel}) holds with the $pv_0$ on the right-hand side negligible compared with $\Delta G^*_p$ and $-\frac{1}{2} Z W(1)$ in this limit. [Cf. the dimensionless  $\pi (=pv_0/\epsilon_{11})$ in~(\ref{eq:solventnumbers})].  Numerically, with $Z=3$, $a=1/3$, $t=1/5$, these asymptotic formulas yield the approximations $y\approx 0.99945$, $\Sigma \approx 0.082$, $D \approx -341$, $\Delta G^*_p/kT \sim -\frac{3}{2}W(1)/kT \approx 2.5$, $B_1/v_0 \approx -13.8$, and $B_2/v_0 \approx -1.3$, which may be compared with their exact values in~(\ref{eq:solventnumbers}) and~(\ref{eq:solutenumbers}).
\section{\label{summary}Summary}

We have studied a lattice model of a solution in which a solute is at infinite dilution in a liquid solvent.  The lattice is a Bethe lattice (Cayley tree) of coordination number $Z$.  The model parameters are ultimately chosen to make the solute solubility $\Sigma$, defined in~(\ref{eq:def:sigma}), low, which then results in a strong, solvent-mediated attraction between solute molecules.

The properties of the solvent alone, which are required in the later parts of the paper, are summarized in Section~\ref{solvent}.  The main new results are the properties of the model solution obtained in Section~\ref{solution} and in the appendices, and the low-temperature asymptotic limits in Section~\ref{limits}.

The solute-solute, solvent-solute, and solvent-solvent pair-correlation functions $h_{22}(r)$, $h_{12}(r)$, and $h_{11}(r)$ are found in~(\ref{eq:relh}) to be related by $h_{22}(r)/h_{12}(r) \equiv h_{12}(r)/h_{11}(r)$ for all $r \geq  2$.  This is remarked to be a special case of an important statistical-mechanical theorem on the asymptotic ($r\rightarrow \infty$) behavior of the pair correlations in mixtures, as discussed by Evans et al.~\cite{evans94:jcp}.  In the low-solubility, high-solvent-density states of interest, the common value $D$ of those two ratios is found to be negative and large in absolute value.  The result is that the amplitude of $h_{22}(r)$ is orders of magnitude greater than that of $h_{11}(r)$, and so is still discernible at distances beyond those at which $h_{11}(r)$ has become negligible.  This is in agreement with what was observed in an earlier model~\cite{widom07:jpcc,widom08:mp}.  This strong effective attraction between pairs of solute molecules reflects the low solubility;\textit{ i.e.}, the positive (thus, unfavorable) constant-pressure solvation free energy $\Delta G^*_p = -kT \ln \Sigma$.  This is seen also in the second osmotic virial coefficient $B$, which in the states of interest is large and negative.  The main contributor to $B$ in these states is $h_{22}(1)$, the solute-solute pair correlation function at contact, which shows the propensity to dimerization.

The effective attraction between solutes that accompanies low solubility is seen also in the quantitative relation~(\ref{eq:lrel}), which relates the solvation free energy $\Delta G^*_p$ to $W(1)$, the solvent-mediated part of the potential of mean force between pairs of solute molecules at contact.  The pressure term $pv_0$ on the right-hand side is usually negligible compared with both $\Delta G^*_p$ and $-\frac{1}{2}Z W(1)$ and may thus often be taken to be $0$.  Near linearity of the relation of         $\Delta G^*_p$   to $W(1)$ was observed in an earlier lattice model as well~\cite{widom03:pccp}, and confirmed in computer simulations of aqueous solutions of non-polar solutes~\cite{paschek04:jcp}. Fitted values of the coordination number $Z$ are then in the range $4.2$ to $5.8$.

In the states of primary interest here, the solubility increases with increasing temperature at fixed pressure.  The entropy of solvation $\Delta S^*_p$ is negative and the 
enthalpy $\Delta H_p$             is positive, so $\Delta S^*_p$  and     $\Delta H_p$     both contribute to the low solubility.  These are in contrast to the hydrophobic effect as seen in aqueous solutions of non-polar solutes, where  $\Sigma$     decreases ($\Delta G^*_p$ increases) with increasing temperature, and $\Delta H_p < 0$, so the solubility is low only because $\Delta S^*_p$ is sufficiently negative to overweigh it.

\section{ACKNOWLEDGMENTS}
We thank Dr.~Paolo De Gregorio for helpful comments and advice. This work was supported by the National Science Foundation.

\appendix

\section{Pair correlation functions}

Here we derive~(\ref{eq:relh}), (\ref{eq:h12c}) and (\ref{eq:h22c}), with a solvent $1$ at density $y$ and a solute $2$ at infinite dilution. We define $P_{\eta \eta'}(r)$ as the probability 
that if a lattice site is occupied by a molecule of species $\eta$, a site $r$ lattice steps from it is occupied by a molecule of species $\eta'$. ``Species'' 0 means unocuppied. For a Cayley tree, with a solute molecule at a specified site, the probability $P_{21}(r)$ of finding a solvent molecule at a site $r$ steps away from it satisfies
\be
P_{21}(r) = P_{21}(r-1)P_{11}(1)+P_{20}(r-1)P_{01}(1). \label{eq:prob1}
\ee
In addition, the $P_{\eta \eta'}(r)$ satisfy $P_{21}(r-1)+P_{20}(r-1)=1$, and the solvent density $y$ at any site is related to the density at its neighbouring sites through:
\be
y = y P_{11}(1)+(1-y)P_{01}(1). \label{eq:prob2}
\ee
$P_{\eta \eta'}(r)$ is related to $h_{\eta \eta'}(r)$ by
\be
P_{\eta \eta'}(r) =  y\left [ 1+h_{\eta \eta'}(r) \right ],
\ee
so from~(\ref{eq:prob1}) and~(\ref{eq:prob2}),
\be
h_{21}(r) = d^{r-1} h_{21}(1),\label{eq:rec1}
\ee
with $h_{21}(1)[=h_{12}(1)]$ from~(\ref{eq:h12c}) and $d= P_{11}(1) - P_{01}(1)$.

These also hold when solute and solvent are identical, so 
\be
h_{11}(r) = d^{r-1} h_{11}(1)\label{eq:rec2},
\ee
from~(\ref{eq:rec1}) and (\ref{eq:rec2}),
\be
\frac{h_{21}(r)}{h_{11}(r)}=\frac{h_{21}(1)}{h_{11}(1)} \equiv D,\label{eq:crossproduct}
\ee
which is the first of the identities~(\ref{eq:relh}).

\begin{figure}
\begin{center}
\includegraphics[scale=0.5]{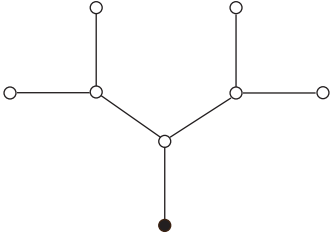}
\caption{A branch of the Cayley tree with coordination number $Z=3$ and a solute on its base site.}
\label{fig:branch}
\end{center}
\end{figure}

Next we calculate the pair distribution function $g_{22}(r)[=1+h_{22}(r)]$, which is given by the potential distribution theorem~\cite{beck:potential}:
\be
g_{22}(r) = c^{b \delta_{1r} }  \frac{ \left \langle  e^{-\beta [ U'_{2,B} (0)+U'_{2,B}(r)]}  \right \rangle  }{ \left \langle  e^{-\beta  U'_{2,B} (0)}
 \right \rangle^2 }, \label{eq:gss}
\ee
where $\beta=1/kT$, $\delta_{ij}$ is the Kronecker delta, and $U'_{2,B}(k)$ is the total interaction energy between a solute molecule and the surrounding fluid. The denominator appearing in~(\ref{eq:gss}) is related to~(\ref{eq:sigma}) through  $\Sigma =  \langle e^{-\beta  U'_{2,B} (0)} \rangle$.
The numerator of~(\ref{eq:gss}), $\langle  e^{-\beta [ U'_{2,B} (0)+U'_{2,B}(r)]} \rangle$, will be calculated using transfer matrices~\cite{Salinas:introduction}.
To simplify the representation of the matrices' elements we will follow ref.~\cite{hu98:pre} and use the so called normalized partial partition functions $x_\eta$, which correspond to the ratio between partition functions of branches with a molecule $\eta$ and a hole on its base site (Fig.~\ref{fig:branch}). 
These functions are related to the quantities previously defined in the text by
\begin{subequations}
\bea
x_1 & = & \frac{y}{(1-y)\alpha}, \label{eq:x1}\\
x_2 & = & \frac{1+\alpha c^a}{1+\alpha}\label{eq:x2}.
\eea
\end{subequations}

The transfer matrices connecting different lattice types are given by:
\bea
\left [ V^{(11)} \right ]_{\eta \eta'} & = &  \left [ V \right ]_{\eta \eta'} \nonumber \\
                                       & = & c^{ \eta \eta' } x_1^{(Z-2)(\eta+\eta')/2}z^{(\eta+\eta')/2}\\
\left [ V^{(21)} \right ]_{\eta \eta'} & = & \left [ V^{(12)} \right ]_{\eta' \eta}  \nonumber \\
                                       & = & \delta_{0\eta}{c^a}^{(1-\eta)\eta'} x_1^{(Z-2)\eta'/2}z^{\eta'/2},
\eea
where $\eta$ and $\eta'$ are occupation variables restricted to unocuppied sites (0) and solvent (1), and where $z$ is the solvent activity, which satisfies $z x_1^Z = y/(1-y)$.

To obtain an expression for $g_{22}(r)$ in terms of a trace, we connect sites $0$ and $r$ with \textit{ghost links} whose elements are:
\begin{subequations}
\bea
\left [ A^{(11)} \right ]_{\eta \eta'} &=& x_1^{Z(\eta + \eta')/2}z^{(\eta + \eta')/2}\label{eq:ghost11} \\
\left [ A^{(22)} \right ]_{\eta \eta'} &=& \delta_{0\eta}\delta_{0\eta'}x_2^{(Z-1)(2-\eta -\eta')}\label{eq:ghost22}.
\eea
\end{subequations}

With the above definitions, and for $r>2$,
\be
 g_{22} (r) =  \dfrac{1}{\Sigma^2}\frac{\texttt{Tr}\{A^{(22)} V^{(21)}V^{r-2} V^{(12)}\} }{\texttt{Tr}\{ A^{(11)} V^{r} \} }. \label{eq:gss2}
\ee
After some algebra, and by using~(\ref{eq:x1}) and (\ref{eq:x2}), one finds:
\be
g_{22}(r) = 1+ \frac{1-y}{y}D^2 e^{-r/\xi},
\label{eq:gss:final}
\ee
where $\xi$ is given by~(\ref{eq:xi}) and
\be
D^2 = \frac{h_{22}(1)}{h_{11}(1)},
\ee
which completes~(\ref{eq:relh}).


Since eq.~(\ref{eq:gss:final}) is valid only for $r\geq2$, it is necessary to calculate $h_{22}(1)$.
We recalculate~(\ref{eq:gss}) for $r=1$ using
\be
\left \langle  e^{-\beta [ U'_{2,B} (0)+U'_{2,B}(1)]} \right \rangle =  \frac{\texttt{Tr}\{A^{(22)} V^{(22)}\} }{\texttt{Tr}\{ A^{(11)} V \} }, \label{eq:caverage}
\ee
with $V^{(22)}$ defined as
\be
\left [ V^{(22)} \right ]_{\eta \eta'}  =   \delta_{0\eta}\delta_{0\eta'} \label{eq:V22}. 
\ee
After some manipulations~(\ref{eq:gss}), at contact, becomes
\be
g_{22}(1) = \frac{c^b}{ x_2^2 } \frac{x_1}{ (1-y)x_1+y }. \label{eq:g22}
\ee
Together with eqs.~(\ref{eq:h11}) and~(\ref{eq:relh}), this completes the evaluation of $h_{12}(r)$ and $h_{22}(r)$.

\section{The relation between $W(1)$ and $\Delta G^*_p$}

The relevant part of the potential of mean force between two solutes does not contain the solute-solute interaction, being exclusively the solvent-mediated part of the effective interaction. At contact, $r=1$, this is
given by
\be
W(1) = -kT \ln \left [ g_{22}(1) \right ] + \epsilon_{22}.
\ee
From equation~(\ref{eq:g22}) it becomes:
\be
 W(1) = 2 kT \ln x_2  + kT \ln \left [ \frac{ (1+y)x_1+y }{x_1}  \right ]. \label{eq:W22a}
\ee
The standard solvation free energy of the solute 2, from (\ref{eq:sigma}), (\ref{eq:def:dG}), and (\ref{eq:x2}), is:
\be
\Delta G^*_p = - kT \ln (1-y) -Z kT \ln x_2, \label{eq:dG2}
\ee
so from eqs.~(\ref{eq:W22a}) and~(\ref{eq:dG2}), 
\be
\frac{Z}{2} W(1) + \Delta G^*_p = kT \ln \left \{ \left [ \frac{ (1+y)x_1+y }{x_1} \right ]^{ \frac{Z}{2}  }  \frac{ 1 }{ 1-y } \right \}. \label{eq:linear1}
\ee

Using eqs.~(\ref{eq:def})-(\ref{eq:Q}) and~(\ref{eq:x1}) one may show that the right hand side of~(\ref{eq:linear1}) is $pv_0$, which then gives~(\ref{eq:lrel}).


\begin{thebibliography}{29}
\expandafter\ifx\csname natexlab\endcsname\relax\def\natexlab#1{#1}\fi
\expandafter\ifx\csname bibnamefont\endcsname\relax
  \def\bibnamefont#1{#1}\fi
\expandafter\ifx\csname bibfnamefont\endcsname\relax
  \def\bibfnamefont#1{#1}\fi
\expandafter\ifx\csname citenamefont\endcsname\relax
  \def\citenamefont#1{#1}\fi
\expandafter\ifx\csname url\endcsname\relax
  \def\url#1{\texttt{#1}}\fi
\expandafter\ifx\csname urlprefix\endcsname\relax\def\urlprefix{URL }\fi
\providecommand{\bibinfo}[2]{#2}
\providecommand{\eprint}[2][]{\url{#2}}

\bibitem[{\citenamefont{Rushbrooke}(1949)}]{rushbrooke:statmech}
\bibinfo{author}{\bibfnamefont{G.~S.} \bibnamefont{Rushbrooke}},
  \emph{\bibinfo{title}{Introduction to Statistical Mechanics}}
  (\bibinfo{publisher}{Oxford University Press}, \bibinfo{year}{1949}),
  \bibinfo{note}{pp. 300-304.}

\bibitem[{\citenamefont{Hill}(1956)}]{hill:statmech}
\bibinfo{author}{\bibfnamefont{T.~L.} \bibnamefont{Hill}},
  \emph{\bibinfo{title}{Statistical Mechanics}}
  (\bibinfo{publisher}{McGraw-Hill}, \bibinfo{year}{1956}), \bibinfo{note}{pp.
  348-353.}

\bibitem[{\citenamefont{Bell and Lavis}(1970{\natexlab{a}})}]{bell70:jphysa1}
\bibinfo{author}{\bibfnamefont{G.~M.} \bibnamefont{Bell}} \bibnamefont{and}
  \bibinfo{author}{\bibfnamefont{D.~A.} \bibnamefont{Lavis}},
  \bibinfo{journal}{J. Phys. A: Gen. Phys.} \textbf{\bibinfo{volume}{3}},
  \bibinfo{pages}{427} (\bibinfo{year}{1970}{\natexlab{a}}).

\bibitem[{\citenamefont{Bell and Lavis}(1970{\natexlab{b}})}]{bell70:jphysa2}
\bibinfo{author}{\bibfnamefont{G.~M.} \bibnamefont{Bell}} \bibnamefont{and}
  \bibinfo{author}{\bibfnamefont{D.~A.} \bibnamefont{Lavis}},
  \bibinfo{journal}{J. Phys. A: Gen. Phys.} \textbf{\bibinfo{volume}{3}},
  \bibinfo{pages}{568} (\bibinfo{year}{1970}{\natexlab{b}}).

\bibitem[{\citenamefont{Besseling and Lyklema}(1997)}]{besseling97}
\bibinfo{author}{\bibfnamefont{N.~A.~M.} \bibnamefont{Besseling}}
  \bibnamefont{and} \bibinfo{author}{\bibfnamefont{J.}~\bibnamefont{Lyklema}},
  \bibinfo{journal}{J. Phys. Chem. B} \textbf{\bibinfo{volume}{101}},
  \bibinfo{pages}{7604} (\bibinfo{year}{1997}).

\bibitem[{\citenamefont{Eads}(2002)}]{eads02}
\bibinfo{author}{\bibfnamefont{C.~D.} \bibnamefont{Eads}}, \bibinfo{journal}{J.
  Phys. Chem. B} \textbf{\bibinfo{volume}{106}}, \bibinfo{pages}{12282}
  (\bibinfo{year}{2002}).

\bibitem[{\citenamefont{Widom et~al.}(2003)\citenamefont{Widom, Bhimalapuram,
  and Koga}}]{widom03:pccp}
\bibinfo{author}{\bibfnamefont{B.}~\bibnamefont{Widom}},
  \bibinfo{author}{\bibfnamefont{P.}~\bibnamefont{Bhimalapuram}},
  \bibnamefont{and} \bibinfo{author}{\bibfnamefont{K.}~\bibnamefont{Koga}},
  \bibinfo{journal}{Phys. Chem. Chem. Phys.} \textbf{\bibinfo{volume}{5}},
  \bibinfo{pages}{3085} (\bibinfo{year}{2003}).

\bibitem[{\citenamefont{Guisoni and Henriques}(2006)}]{henriques06}
\bibinfo{author}{\bibfnamefont{N.}~\bibnamefont{Guisoni}} \bibnamefont{and}
  \bibinfo{author}{\bibfnamefont{V.~B.} \bibnamefont{Henriques}},
  \bibinfo{journal}{J. Phys. Chem. B} \textbf{\bibinfo{volume}{110}},
  \bibinfo{pages}{17188} (\bibinfo{year}{2006}).

\bibitem[{\citenamefont{Beck et~al.}(2006)\citenamefont{Beck, Paulaitis, and
  Pratt}}]{beck:potential}
\bibinfo{author}{\bibfnamefont{T.~L.} \bibnamefont{Beck}},
  \bibinfo{author}{\bibfnamefont{M.~E.} \bibnamefont{Paulaitis}},
  \bibnamefont{and} \bibinfo{author}{\bibfnamefont{L.~R.} \bibnamefont{Pratt}},
  \emph{\bibinfo{title}{The Potential Distribution Theorem and Models of
  Molecular Solutions}} (\bibinfo{publisher}{Cambridge University Press},
  \bibinfo{year}{2006}), \bibinfo{note}{{C}hap. 7.}

\bibitem[{\citenamefont{Rogers and Beck}(2008)}]{rogers08}
\bibinfo{author}{\bibfnamefont{D.~M.} \bibnamefont{Rogers}} \bibnamefont{and}
  \bibinfo{author}{\bibfnamefont{T.~L.} \bibnamefont{Beck}},
  \bibinfo{journal}{J. Chem. Phys.} \textbf{\bibinfo{volume}{129}},
  \bibinfo{pages}{134505} (\bibinfo{year}{2008}).

\bibitem[{\citenamefont{Barbosa and Henriques}(2008)}]{barbosa08}
\bibinfo{author}{\bibfnamefont{M.~A.~A.} \bibnamefont{Barbosa}}
  \bibnamefont{and} \bibinfo{author}{\bibfnamefont{V.~B.}
  \bibnamefont{Henriques}}, \bibinfo{journal}{Phys. Rev. E}
  \textbf{\bibinfo{volume}{77}}, \bibinfo{pages}{051204}
  (\bibinfo{year}{2008}).

\bibitem[{\citenamefont{Buzano and Pretti}(2003)}]{buzano03:jcp}
\bibinfo{author}{\bibfnamefont{C.}~\bibnamefont{Buzano}} \bibnamefont{and}
  \bibinfo{author}{\bibfnamefont{M.}~\bibnamefont{Pretti}},
  \bibinfo{journal}{J. Chem. Phys.} \textbf{\bibinfo{volume}{119}},
  \bibinfo{pages}{3791} (\bibinfo{year}{2003}).

\bibitem[{\citenamefont{Wheeler and Widom}(1970)}]{wheeler70:jcp}
\bibinfo{author}{\bibfnamefont{J.~C.} \bibnamefont{Wheeler}} \bibnamefont{and}
  \bibinfo{author}{\bibfnamefont{B.}~\bibnamefont{Widom}}, \bibinfo{journal}{J.
  Chem. Phys.} \textbf{\bibinfo{volume}{52}}, \bibinfo{pages}{5334}
  (\bibinfo{year}{1970}).

\bibitem[{\citenamefont{Kirkwood and Buff}(1951)}]{kirkwood51}
\bibinfo{author}{\bibfnamefont{J.~G.} \bibnamefont{Kirkwood}} \bibnamefont{and}
  \bibinfo{author}{\bibfnamefont{F.~P.} \bibnamefont{Buff}},
  \bibinfo{journal}{J. Chem. Phys.} \textbf{\bibinfo{volume}{19}},
  \bibinfo{pages}{774} (\bibinfo{year}{1951}).

\bibitem[{\citenamefont{Zimm}(1953)}]{zimm53}
\bibinfo{author}{\bibfnamefont{B.~H.} \bibnamefont{Zimm}}, \bibinfo{journal}{J.
  Chem. Phys.} \textbf{\bibinfo{volume}{21}}, \bibinfo{pages}{934}
  (\bibinfo{year}{1953}).

\bibitem[{\citenamefont{Ben-Naim}(1974)}]{ben-naim:water}
\bibinfo{author}{\bibfnamefont{A.}~\bibnamefont{Ben-Naim}},
  \emph{\bibinfo{title}{Water and Aqueous Solutions: Introduction to a
  Molecular Theory}} (\bibinfo{publisher}{Plenum Press}, \bibinfo{year}{1974}),
  \bibinfo{note}{p. 142.}

\bibitem[{\citenamefont{Ben-Naim}(1977)}]{ben-naim77:jcp}
\bibinfo{author}{\bibfnamefont{A.}~\bibnamefont{Ben-Naim}},
  \bibinfo{journal}{J. Chem. Phys.} \textbf{\bibinfo{volume}{67}},
  \bibinfo{pages}{4884} (\bibinfo{year}{1977}).

\bibitem[{\citenamefont{{De Gregorio} and Widom}(2007)}]{widom07:jpcc}
\bibinfo{author}{\bibfnamefont{P.}~\bibnamefont{{De Gregorio}}}
  \bibnamefont{and} \bibinfo{author}{\bibfnamefont{B.}~\bibnamefont{Widom}},
  \bibinfo{journal}{J. Phys. Chem. C} \textbf{\bibinfo{volume}{111}},
  \bibinfo{pages}{16060} (\bibinfo{year}{2007}).

\bibitem[{\citenamefont{Evans et~al.}(1994)\citenamefont{Evans, {Leote de
  Carvalho}, Henderson, and Hoyle}}]{evans94:jcp}
\bibinfo{author}{\bibfnamefont{R.}~\bibnamefont{Evans}},
  \bibinfo{author}{\bibfnamefont{R.~J.~F.} \bibnamefont{{Leote de Carvalho}}},
  \bibinfo{author}{\bibfnamefont{J.~R.} \bibnamefont{Henderson}},
  \bibnamefont{and} \bibinfo{author}{\bibfnamefont{D.~C.} \bibnamefont{Hoyle}},
  \bibinfo{journal}{J. Chem. Phys.} \textbf{\bibinfo{volume}{100}},
  \bibinfo{pages}{591} (\bibinfo{year}{1994}).

\bibitem[{\citenamefont{{De Gregorio} et~al.}(2008)\citenamefont{{De Gregorio},
  Toledo, and Widom}}]{widom08:mp}
\bibinfo{author}{\bibfnamefont{P.}~\bibnamefont{{De Gregorio}}},
  \bibinfo{author}{\bibfnamefont{J.~C.} \bibnamefont{Toledo}},
  \bibnamefont{and} \bibinfo{author}{\bibfnamefont{B.}~\bibnamefont{Widom}},
  \bibinfo{journal}{Mol. Phys.} \textbf{\bibinfo{volume}{106}},
  \bibinfo{pages}{419} (\bibinfo{year}{2008}).

\bibitem[{\citenamefont{Tucker and Christian}(1979)}]{tucker79:jpc}
\bibinfo{author}{\bibfnamefont{E.~E.} \bibnamefont{Tucker}} \bibnamefont{and}
  \bibinfo{author}{\bibfnamefont{S.~D.} \bibnamefont{Christian}},
  \bibinfo{journal}{J. Phys. Chem.} \textbf{\bibinfo{volume}{83}},
  \bibinfo{pages}{426} (\bibinfo{year}{1979}).

\bibitem[{\citenamefont{Rossky and Friedman}(1980)}]{rossky80:jpc}
\bibinfo{author}{\bibfnamefont{P.~J.} \bibnamefont{Rossky}} \bibnamefont{and}
  \bibinfo{author}{\bibfnamefont{H.~L.} \bibnamefont{Friedman}},
  \bibinfo{journal}{J. Phys. Chem.} \textbf{\bibinfo{volume}{84}},
  \bibinfo{pages}{587} (\bibinfo{year}{1980}).

\bibitem[{\citenamefont{Chipot et~al.}(1996)\citenamefont{Chipot, Kollman, and
  Pearlman}}]{chipot96:jcc}
\bibinfo{author}{\bibfnamefont{C.}~\bibnamefont{Chipot}},
  \bibinfo{author}{\bibfnamefont{P.~A.} \bibnamefont{Kollman}},
  \bibnamefont{and} \bibinfo{author}{\bibfnamefont{D.~A.}
  \bibnamefont{Pearlman}}, \bibinfo{journal}{J. Computational Chem.}
  \textbf{\bibinfo{volume}{17}}, \bibinfo{pages}{1112} (\bibinfo{year}{1996}).

\bibitem[{\citenamefont{Liu and Ruckenstein}(1998)}]{liu98:jpcb}
\bibinfo{author}{\bibfnamefont{H.}~\bibnamefont{Liu}} \bibnamefont{and}
  \bibinfo{author}{\bibfnamefont{E.}~\bibnamefont{Ruckenstein}},
  \bibinfo{journal}{J. Phys. Chem. B} \textbf{\bibinfo{volume}{102}},
  \bibinfo{pages}{1005} (\bibinfo{year}{1998}).

\bibitem[{\citenamefont{Coskuner and Deiters}(2006)}]{coskuner06:zpc}
\bibinfo{author}{\bibfnamefont{O.}~\bibnamefont{Coskuner}} \bibnamefont{and}
  \bibinfo{author}{\bibfnamefont{U.~K.} \bibnamefont{Deiters}},
  \bibinfo{journal}{Z. Phys. Chem.} \textbf{\bibinfo{volume}{220}},
  \bibinfo{pages}{349} (\bibinfo{year}{2006}).

\bibitem[{\citenamefont{Coskuner and Deiters}(2007)}]{coskuner07:zpc}
\bibinfo{author}{\bibfnamefont{O.}~\bibnamefont{Coskuner}} \bibnamefont{and}
  \bibinfo{author}{\bibfnamefont{U.~K.} \bibnamefont{Deiters}},
  \bibinfo{journal}{Z. Phys. Chem.} \textbf{\bibinfo{volume}{221}},
  \bibinfo{pages}{785} (\bibinfo{year}{2007}).

\bibitem[{\citenamefont{Paschek}(2004)}]{paschek04:jcp}
\bibinfo{author}{\bibfnamefont{D.}~\bibnamefont{Paschek}}, \bibinfo{journal}{J.
  Chem. Phys.} \textbf{\bibinfo{volume}{120}}, \bibinfo{pages}{6674}
  (\bibinfo{year}{2004}).

\bibitem[{\citenamefont{Salinas}(2001)}]{Salinas:introduction}
\bibinfo{author}{\bibfnamefont{S.~R.~A.} \bibnamefont{Salinas}},
  \emph{\bibinfo{title}{Introduction to statistical physics}}
  (\bibinfo{publisher}{Springer-Verlag}, \bibinfo{address}{New York},
  \bibinfo{year}{2001}), \bibinfo{note}{{C}hap. 13.}

\bibitem[{\citenamefont{Hu and Izmailian}(1998)}]{hu98:pre}
\bibinfo{author}{\bibfnamefont{C.-K.} \bibnamefont{Hu}} \bibnamefont{and}
  \bibinfo{author}{\bibfnamefont{N.~S.} \bibnamefont{Izmailian}},
  \bibinfo{journal}{Phys. Rev. E} \textbf{\bibinfo{volume}{58}},
  \bibinfo{pages}{1644} (\bibinfo{year}{1998}).

\end{thebibliography}

\end{document}